\documentclass[10pt,letterpaper]{article}
\usepackage[top=0.85in,left=2.75in,footskip=0.75in]{geometry}
\usepackage{amsmath,amssymb}
\usepackage{changepage}
\usepackage[utf8x]{inputenc}
\usepackage{textcomp,marvosym}
\usepackage{cite}
\usepackage{nameref,hyperref}
\usepackage[right]{lineno}
\usepackage{microtype}
\DisableLigatures[f]{encoding = *, family = * }
\usepackage[table]{xcolor}
\usepackage{array,multirow}
\usepackage{booktabs}
\usepackage{color}

\newcolumntype{+}{!{\vrule width 2pt}}

\newlength\savedwidth

\raggedright
\usepackage[aboveskip=1pt,labelfont=bf,labelsep=period,justification=raggedright,singlelinecheck=off]{caption}

\makeatletter
\renewcommand{\@biblabel}[1]{\quad#1.}
\makeatother

\date{}

\usepackage{lastpage,fancyhdr,graphicx}
\usepackage{epstopdf}
\fancyheadoffset[L]{2.25in}
\fancyfootoffset[L]{2.25in}
\newcommand{\ffrac}[2]{\ensuremath{\frac{\displaystyle #1}{\displaystyle #2}}}
\begin{document}
\vspace*{0.2in}

\begin{flushleft}
{\Large
\textbf\newline{Theoretical open-loop model of respiratory mechanics in the extremely preterm infant} % Please use "sentence case" for title and headings (capitalize only the first word in a title (or heading), the first word in a subtitle (or subheading), and any proper nouns).
}
\newline
\\
Laura Ellwein Fix\textsuperscript{1*},
Joseph Khoury\textsuperscript{2},
Russell R Moores, Jr.\textsuperscript{2},
Lauren Linkous\textsuperscript{1},
Matthew Brandes\textsuperscript{3},
Henry J. Rozycki\textsuperscript{2}
\\
\bigskip
\textbf{1} Department of Mathematics and Applied Mathematics, Virginia Commonwealth University, Richmond, VA
\\
\textbf{2} Division of Neonatal Medicine, Children's Hospital of Richmond, Virginia Commonwealth University, Richmond, VA
\\
\textbf{3} VCU School of Medicine, Virginia Commonwealth University, Richmond, VA
\bigskip

% Insert additional author notes using the symbols described below. Insert symbol callouts after author names as necessary.
% 
% Remove or comment out the author notes below if they aren't used.
%

% Use the asterisk to denote corresponding authorship and provide email address in note below.
* lellwein@vcu.edu

\end{flushleft}
\section*{Abstract}
Non-invasive ventilation is increasingly used for respiratory support in preterm infants, and is associated with a lower risk of chronic lung disease. However, this mode is often not successful in the extremely preterm infant in part due to their markedly increased chest wall compliance that does not provide enough structure against which the forces of inhalation can generate sufficient pressure. To address the continued challenge of studying treatments in this fragile population, we developed a nonlinear lumped-parameter model of respiratory system mechanics of the extremely preterm infant that incorporates nonlinear lung and chest wall compliances and lung volume parameters tuned to this population. In particular we developed a novel empirical representation of progressive volume loss based on compensatory alveolar pressure increase resulting from collapsed alveoli. The model demonstrates increased rate of volume loss related to high chest wall compliance, and simulates laryngeal braking for elevation of end-expiratory lung volume and constant positive airway pressure (CPAP). The model predicts that low chest wall compliance (chest stiffening) in addition to laryngeal braking and CPAP enhance breathing and delay lung volume loss. These results motivate future data collection strategies and investigation into treatments for chest wall stiffening.

% Please keep the Author Summary between 150 and 200 words
% Use first person. PLOS ONE authors please skip this step. 
% Author Summary not valid for PLOS ONE submissions.   

% Use "Eq" instead of "Equation" for equation citations.
\section*{Introduction}
The extremely  preterm infant, born at $<28$ weeks gestation and usually $<1000$g, is at risk of developing chronic lung disease despite established treatments such as surfactant replacement therapy. Currently the survival rate of this group ranges from $94\%$ at 27 weeks to as low as ~$33\%$ at 23 weeks~\cite{Stoll15}, with survivors living with varying degrees of morbidity. One risk factor for lung disease remains the trauma associated with traditional mechanical ventilation including endotracheal tube injury, high cyclic tidal volumes pressures, and hyperoxia. Non-invasive methods of ventilation such as continuous positive airway pressure (CPAP) are being used with more frequency and have been successful with more mature infants but appear to fail in the extremely preterm infant~\cite{Manley13,Bhandari13,Siew15}. One hypothesis for the failure of non-invasive ventilation and the need for increasing invasive respiratory support is the markedly increased compliance (floppiness) of the chest wall in the extremely preterm infant resulting from ribcage undermineralization common at the start of the third trimester~\cite{Love53,Beltrand08,Kovacs15}. In the preterm infant chest wall compliance can be up to five times lung tissue compliance~\cite{Gerhardt80}. 

When the chest wall is not sufficiently rigid, the negative pressure within the pleural space between the lung and chest wall generated from diaphragm contraction is diminished~\cite{Mortola83}. In many cases this leads to progressive lung collapse (atelectasis) with each breath as the forces needed to open airspaces after each exhalation become insurmountable~\cite{Frappell05}, leading to decreasing lung compliance and functional residual capacity (FRC)~\cite{Miller05}. This progression of events is observed clinically in X-rays and by symptoms of respiratory distress such as chest retractions and rapid breathing. The clinical result is progressively reduced tidal volumes and end-expiratory lung volume (EELV) as the forces needed to open airspaces after exhalation are insufficient. Non-invasive ventilation has been observed to be become ineffective under these conditions, necessitating placement of an endotracheal tube and positive pressure mechanical ventilation and markedly increasing the risk of lung damage.

Despite this being repeatedly observed clinically, there remains little quantification of the impact of variable nonlinear chest wall compliance on tidal breathing dynamics, and even fewer computational modeling efforts investigating the underlying mechanics of progressive volume loss. Most computational models of breathing address the extremes of lung capacity such as a forced vital capacity maneuver, study a static, excised, or injured lung, or use an animal model~\cite{Liu98,Narusawa01,Athan00,Frazer13,Uzawa15}. Existing computer models of tidal breathing have not fully accounted for the physiology particular to premature infants and thus have limited applicability. Often, methods of providing ventilator support have been developed in adults and children, then refined and scaled for newborns and premature infants, limiting innovation aimed specifically at this vulnerable population. 

In this work, we have developed a nonlinear computational model of respiratory mechanics parameterized for the extremely preterm infant that demonstrates differential volume loss under high vs low chest wall compliance conditions. We adapt a model first presented by Athanasiades et al~\cite{Athan00} and modified for newborn lambs by LeRolle et al~\cite{LeRolle13}. In the latter, differences such as smaller diameter airways, higher respiratory rates, higher lung resistance, and higher chest wall compliances were considered, however many of the critical physiological nonlinearities contributing to long-term dynamics were not included. The present model is built upon the nonlinear compliance curves describing pressure-volume relationships specific to preterm infants~\cite{Smith76}. Dynamic alterations of compliance curves based on breath-to-breath end-inspiratory lung volume (EILV) and peak inspiratory pressure (PIP) are shown to influence tidal volume and EELV, thus simulating progressive lung volume loss. We also demonstrate the effect of two simulated interventions that raise alveolar pressure and lung elastic recoil: CPAP which raises the pressure at the mouth; and laryngeal braking (grunting), which increases upper airway resistance during expiration.

%%% TABLE 1
\begin{table}[!ht]
%\centering
\caption{Glossary}
{\small
 \begin{tabular}{ll} 
\toprule
	Parameter/State 								& Physiologic description									  							\\ 
\noalign {\smallskip} \hline \hline
\noalign {\smallskip}
TLC [ml]									& Total lung capacity 																\\	
RV [ml]										& Residual volume  																	\\	
FRC [ml] 									& Functional residual capacity 										\\	
VC[ml]										& Vital capacity 																	\\	
RR [br/min] 							& Respiratory rate																\\	
$f$ [br/s]								& Respiratory frequency																\\	
$T$ [s]										& Duration of respiratory cycle												\\	
$V_T$ [ml]								& Tidal volume																	 \\		
$\dot{V}_E$ [ml/min]			& Minute ventilation															\\	
$\dot{V}_A$ [ml/s]				& Airflow																					\\	
$A_{mus}$ [cm H$_2$O]			& Muscle pressure amplitude												\\	
$P_{tm}$ [cm H$_2$O]				& Transmural pressure															\\	
$P_{A}$ [cm H$_2$O]				& Alveolar pressure																\\	
$P_{el}$ [cm H$_2$O]			& Lung elastic recoil (transpulmonary pressure)		\\	
$P_{ve}$ [cm H$_2$O]			& Viscoelastic component of pressure									\\
$P_{l,dyn}$ [cm H$_2$O]		& Dynamic pulmonary pressure															\\
$P_{pl}$ [cm H$_2$O]			& Pleural pressure																\\	
$P_{cw}$ [cm H$_2$O]			& Chest wall elastic recoil																\\	
$P_{mus}$ [cm H$_2$O]			& Respiratory muscle pressure												\\	
$C_A$ [ml/cm H$_2$O]			& Lung compliance																	\\	
$C_w$ [ml/cm H$_2$O]			& Chest wall compliance														\\	
$C_{rs}$ [ml/cm H$_2$O]		& Respiratory system compliance										\\	
$R_{rs}$ [cm H$_2$O s/L] 	& Respiratory system resistance										\\	
$\nu$								& Fraction of VC for chest wall relaxation volume 		\\	
$V_0$ [ml] 								& Chest wall relaxation volume												\\	
$\beta$										&	Baseline fraction of lung recruited at $P_{el}=0$	 \\
$\gamma$									&	Maximum recruitable function of lung							 \\
$\alpha$									&	Lower asymptote, fraction recruitment							  \\
$k$ [1/cm H$_2$O]					&	Characterizes slope, aggregate lung elasticity		 \\
$c_F$	[cm H$_2$O]					&	Pressure at maximum lung recruitment							  \\
$d_F$	[cm H$_2$O]					&	Characterizes slope at maximum lung recruitment		 \\					
$a_w$ [ml]								& Lower asymptote, chest wall compliance							\\	
$b_w$ [ml]								& Characterizes slope, $P_{cw}\rightarrow \infty$			\\	
$c_w$ [cm H$_2$O]					& Transition point, chest wall compliance							\\	
$d_w$ [cm H$_2$O]					& Characterizes slope, $P_{cw}\rightarrow \infty$		\\	
$a_c$ [ml]								& Lower asymptote, collapsible airway									\\	
$b_c$ [ml]								& Upper asymptote, collapsible airway										\\
$c_c$ [cm H$_2$O]					& Pressure at peak collapsible airway compliance			\\	
$d_c$ [cm H$_2$O]					& Characterizes slope, peak coll. airway compliance	\\
$K_c$ 										& Collapsible airway resistance coefficient					\\	
$V_{c,max}$ [ml]					& Peak collapsible airway volume											\\	
$R_{s,m}$ [cm H$_2$O s/L]	& Minimum small airway resistance											\\	
$R_{s,d}$ [cm H$_2$O s/L]	& Change in small airway resistance										\\	
$K_s$ 										& Small airway resistance low pressure coefficient		\\	
$I_u$ [cm H$_2$O s$^2$/L]	& Upper airway inertance														\\		
$R_{u,m}$ [cm H$_2$O s/L]		& Laminar value, upper airway resistance						\\	
$K_u$ [cm H$_2$O s/L]			& Turbulent coefficient, upper airway resistance		\\	
$C_{ve}$ [L / cm H$_2$O] 	& Lung viscoelastic compliance														\\		
$R_{ve}$ [cm H$_2$O s/L] 	& Lung viscoelastic resistance												\\	
\bottomrule
\end{tabular}}
\label{tab:glossary}
\end{table}

\section*{Mathematical model}

\begin{figure}[!ht]
\includegraphics[scale=0.8]{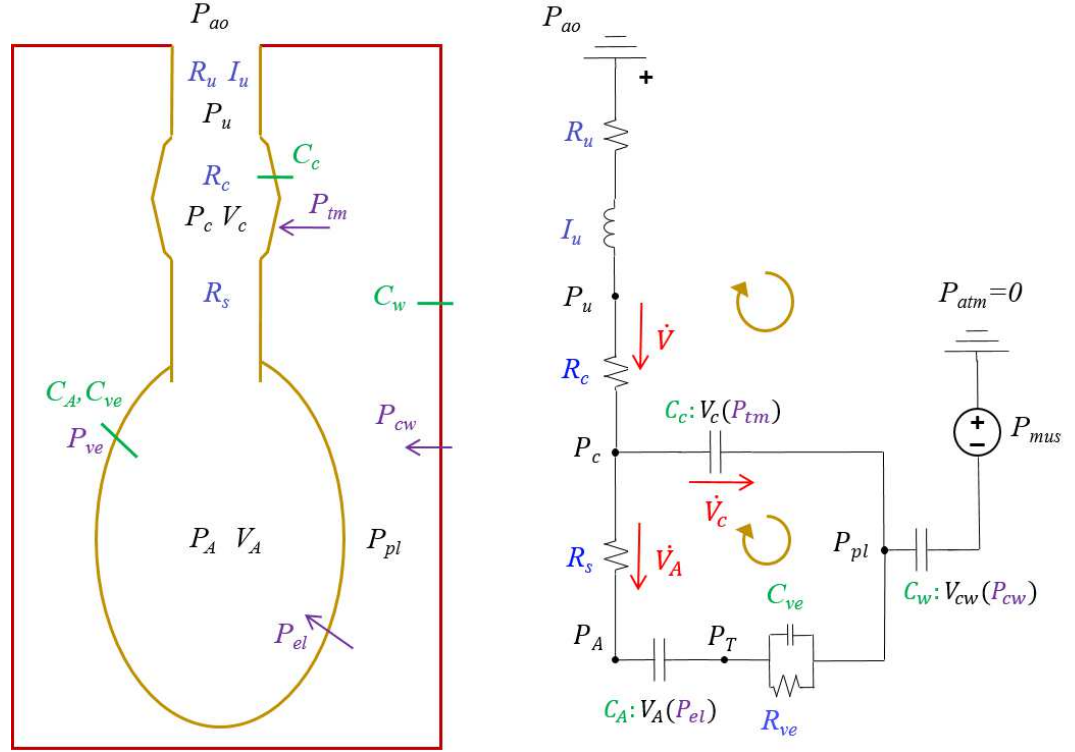}
\caption{{\small {\bf Lumped-parameter respiratory mechanics model, in both volume-pressure (panel A) and electrical (panel B) system analogs.} Each non-rigid compartment has a volume $V$ (black), pressure $P$, (black) and associated compliance $C$ (green, for emphasis) that is a function of the transmural pressures (purple) across the compartment boundaries. Air flows $\dot{V}$ (red) across resistances $R$ and inertance $I$ (blue) are positive in the direction of the arrows. Circular yellow arrows indication direction of loop summations in Eq~(\ref{eq:Pcomp}). Subscripts: airway opening $ao$, upper $u$, collapsible $c$, small peripheral $s$, alveolar $A$, viscoelastic $ve$, lung elastic $el$, transmural $tm$, pleural $pl$, chest wall $cw$, muscle $mus$.}}
\label{fig:model}
\end{figure}

The lumped-parameter respiratory mechanics model describes dynamic volumes and pressures in the airways, lungs, chest wall, and intrapleural space between lungs and chest. A signal that represents diaphragm pressure generated during spontaneous breathing drives the model. A compartment is assumed to display aggregate behavior, e.g. the alveolar compartment represents the collective dynamics of the alveoli as a whole. The model is designed using the volume-pressure analog of an electrical circuit, see Fig~\ref{fig:model}. As such, relevant states are in terms of pressure $P(t)$ [cm H$_2$O] and volume $V(t)$ [ml] in and between air compartments, with volumetric flow rate and rate of change represented as $\dot{V}(t)$ [ml/s] and $\frac{dV}{dt}$ respectively. Air pressure $P_i$ within a specific volume $i$ is defined as the difference between intra-airway pressure $P_{int}$ and pressure external to the body $P_{ext}$, i.e. $P_i=P_{int,i}-P_{ext,i}$. Since all pressures are relative to the same constant atmospheric pressure, all $P_{ext}=0$ and all intra-airway pressures $P_i=P_{int,i}$. The pressure $P_{ij}=P_i-P_j$ refers to the transmural pressure across a compliant boundary separating volumes $i$ and $j$.

%%% FIGURE 1

\subsection*{State equations}

Each non-rigid compartment has an associated compliance $C_i$ [ml/cm H$_2$O], describing the change in compartmental volume $V_i$ given a change in transmural pressure $P_{ij}$ across its boundary with compartment $j$:
\begin{equation*}
C_i=\dfrac{dV_i}{dP_{ij}},
\label{eq:compliance}
\end{equation*} 
The nature of $C_i$ does not change explicitly with time but instead is implicitly determined by the relationship between volume and pressure. This can be reformulated in terms of dynamic changes of state:
\begin{equation*}
\frac{dV_{i}}{dt}=C_i\left(\frac{dP_{ij}}{dt}\right).
\label{eq:volume}
\end{equation*} 

Bidirectional airflow through the trachea, bronchi, bronchioles, and to and from the lungs results from contraction and relaxation of the diaphragm generating a pressure difference. Airflow is opposed by the resistance of the airways as functions of their radaii or tissue properties.  This relationship is described by the flow-pressure analog of Ohm's law~\cite{Mead61}, 
\begin{equation}
\dot{V}_i=\dfrac{P_{i-1}-P_i}{R_i},
\label{eq:Ohms}
\end{equation} 
where $R_i$ [cm H$_2$O$\cdot{}$s/ml] is the resistance to airflow prior to compartment $i$. If a compartment includes inertial effects, the pressure gradient is also a function of the acceleration of flow,
\begin{equation}
P_{i-1}-P_i=I_i\ddot{V}_{i}(t)
\label{eq:inertance}
\end{equation} 
where $I$ is the inertance. Inertial effects are considered for the newborn upper rigid airway because of its smaller radius, but neglected for the rest of the model tissues~\cite{LeRolle13}. 

The pressures $P_{ij}$ across each compliant compartment include transmural pressure between the compliant airways and the pleural space $P_{tm}=P_c-P_{pl}$, lung elastic recoil $P_{el}=P_A-P_T$, lung viscoelastic component $P_{ve}=P_T-P_{pl}$, and chest wall elastic recoil $P_{cw}=P_{pl}-P_{mus}$. Summing pressures over each of three loops according to Kirchhoff's mesh rule gives a system of time-varying algebraic equations:
\begin{eqnarray*}
0&=&P_{ao}-P_c+P_{tm}+P_{pl}\\
0&=&(P_c-P_A)+P_{el}+P_{ve}-P_{tm}\\
0&=&P_{ve}-R_{ve}(\dot{V}_A-\dot{V}_{ve})
\end{eqnarray*}
Using the additional relationships obtained from applying Eq~(\ref{eq:Ohms}-\ref{eq:inertance}), the system of loop equations can be rewritten as 
\begin{eqnarray}
0&=&P_{ao}+R_c\dot{V}-R_u\dot{V}-I_u\ddot{V}+P_{tm}+P_{cw}-P_{mus} \label{eq:Pcomp}\\
0&=&R_s\dot{V}_A+P_{el}+P_{ve}-P_{tm} \nonumber\\
0&=&P_{ve}-R_{ve}(\dot{V}_A-\dot{V}_{ve}). \nonumber
\end{eqnarray}

Rearranging Eq~(\ref{eq:Pcomp}) and using Kirchhoff's current law along with Eq~(\ref{eq:Ohms}-\ref{eq:inertance}) produces the consolidated set of model differential equations:
\begin{eqnarray}
\ddot{V}&:&\frac{d\dot{V}}{dt}=\frac{1}{I_u}\left(P_{ao}-P_u-R_u\dot{V}\right) \label{eq:states}\\
\dot{V}_c&:&\frac{dV_c}{dt}=\dot{V}-\dot{V}_A \nonumber\\
\dot{P}_{el}&:&\frac{dP_{el}}{dt}=\frac{\dot{V}_A}{C_A} \nonumber\\
\dot{P}_{ve}&:&\frac{dP_{ve}}{dt}=\frac{\dot{V}_A-(P_{ve}/R_{ve})}{C_{ve}} \nonumber
\end{eqnarray}

Conservation laws also maintain that $V=V_{cw}=V_A+V_c$, in other words the total system volume equals the chest wall volume, which is the sum of the alveolar and compressible airway volumes. Pressure-volume relationships and compliances $C_i$ will be further described below. 

\subsection*{Nonlinear resistance constitutive relations}

The airways begin with an upper rigid segment characterized by an inertance $I_u$ and a nonlinear Rohrer resistance $R_u$~\cite{LeRolle13,Rohrer15} that increases with airflow:
\begin{equation} 
R_u=R_{u,m}+K_u|\dot{V}| 
\label{eq:Ru}
\end{equation} 
The constants $R_{u,m}$ and $K_u$ represent laminar and turbulent flow components. 

A middle collapsible portion is modeled as a cylinder with constant length having nonlinear resistance $R_c$ that depends inversely on the 4th power of the radius according to Poiseuille's law. Therefore $R_c$ is formulated as~\cite{Olender76,Liu98}:
\begin{equation} 
R_c=K_c\left(\frac{V_{c,max}}{V_c}\right)^2
\label{eq:Rc}
\end{equation} 
where $R_c$ equals its minimum value $K_c$ when $V_c=V_{c,max}$, an estimate of dead space. 
%Note that low $V_c$ corresponds to high $R_c$ and may reflect airway collapse. 

An inverse relationship between resistance in the smaller peripheral airways $R_s$ and lung volume $V_A$ reflects high resistance at low or near-zero volumes~\cite{Olender76}. To avoid $R_s\rightarrow \infty$ as $V_A\rightarrow 0$~\cite{Avanz01} from a strict exponential decay model, we adopt the formulation used by both Liu et al and Athanasiades et al~\cite{Liu98,Athan00}, a decaying exponential function of relative lung volume with finite $R_s$ at $V_A=0$:
\begin{sloppypar}
\begin{equation}
R_s=R_{s,d}\cdot e^{K_s(V_A-RV)/(TLC-RV)}+R_{s,m} 
\label{eq:Rs}
\end{equation} 
where $K_s<0$. This parameterization gives that ${R_s\approx R_{s,m}}$ when $V_A=TLC$, and $R_s=R_{s,d}+R_{s,m}$ when $V_A=RV$ (residual volume). 
\end{sloppypar}

\subsection*{Nonlinear compliance constitutive relations}\label{sec:compliances}

The volumes $V_c$, $V_A$, and $V_{cw}$ representing physiological compartments are assumed to have nonlinear compliance which are modeled implicitly with a pressure-volume curve or explicitly by $C_i=\frac{dV_i}{dP_i}$.

The compliance curve for the collapsible airway volume $V_c$ as a function of $P_{tm}$ represents data depicted in~\cite{Olender76} following a sigmoidal function~\cite{LeRolle13, Avanz01} 
\begin{equation}
V_c=\frac{V_{c,max}}{1+e^{-(P_{tm}-c_c)/d_c}} 
\label{eq:Vc} 
\end{equation}
Maximal compliance occurs at the middle of the sigmoid $c_c$, with $d_c$ characterizing the slope of the sigmoid. 

In newborns and infants an exponential-like chest-wall compliance curve is observed~\cite{Donn98,Goldsmith11} but with compliance being near infinite for $P_{cw}>0$. We chose to model the static compliance of the chest wall as a ``softplus'' function of the form $f(x)=\ln(1+e^x)$, the smooth approximation of the rectifier activation function $f(x)=\max(0,x)$. Accounting for translations and scaling, this is represented by
\begin{equation}
V_{cw}=RV+b_w \ln \left(1+e^{(P_{cw})/d_w}\right)
\label{eq:Vcw}
\end{equation}
The asymptotic volume at large negative pressure is thus assumed to equal RV. The ``transition point'' where the softplus function slope has the greatest rate of change from horizontal to affine occurs at $P_{cw}=$. The chest wall relaxation volume $V_0=V_{cw}|_{P_{cw}=0}$ is set using an estimate from literature at 25\% of VC (vital capacity)~\cite{Donn98,Goldsmith11}. From this parameterization, $b_w=(V_0-RV)/(\ln 2)$. The single degree of freedom $d_w$ then characterizes the slope of the chest wall compliance curve and is adjusted to produce a range of dynamic compliance values.

The volume of the lung compartment $V_A$ is modeled as the product of distention of lung units $V_{el}(P_{el})$ and fraction of recruited alveoli $F_{rec}(P_{el})$~\cite{Venegas94,Hamlington16}. To obtain $V_A\approx RV$ near $P_{el}=0$, lung volume is given as 
\begin{equation}
V_A=V_{el}(P_{el})\cdot F_{rec}(P_{el})+RV.
\label{eq:VA}
\end{equation}
Alveolar compliance $C_A$ as used in the system of differential equations~\eqref{eq:states} is found with symbolic computation as $\frac{dV_A}{dP_{el}}$. 

The first term $V_{el}$ represents the volume due to aggregate elasticity of the lung unit structure, which is modeled here as a saturated exponential ~\cite{Hamlington16,Colebatch79,Ferreira11}
\begin{equation} 
V_{el}=VC\cdot(1-e^{(-kP_{el})})
\label{eq:Vel}
\end{equation}
 where $k$ characterizes the lung stiffness. This representation has been found to suffice in cases of a healthy or surfactant-treated lung. The second term of the lung compliance $F_{rec}$ represents the contribution of recruitment and derecruitment of alveoli to compliance, which has been modeled previously as dependent on both time and pressure~\cite{Venegas94, Hamlington16,Bates02}. It can be represented by a sigmoid which resembles the probability density function of a Gaussian distribution describing aggregate opening or closing pressures of individual alveolar sacs or ducts~\cite{Sundaresan09}. We adopt the formulation of Hamlington et al~\cite{Hamlington16}:
\begin{eqnarray}
F_{rec}&=&\alpha+\frac{\gamma-\alpha}{1+e^{-(P_{el}-c_F)/d_F}}, \label{eq:Frec}\\
\mbox{where} \ \ \ \alpha&=&\frac{(1+e^{c_F/d_F})\beta-\gamma}{e^{c_F/d_F}}. \nonumber
\end{eqnarray}

It follows that $\beta$ is the baseline fraction of lung recruited at $P_{el}=0$, $\gamma$ represents the maximum recruitable fraction of lung, $c_F$ is mean opening pressure at which recruitment is maximum, and $d_F$ describes the transition to full recruitment capturing the heterogeneity of the lung. Parameterization of $F_{rec}$ is based on the state of health being modeled and can change breath-to-breath depending on conditions. For example, an increase in stiffness resulting from derecruitment may manifest as higher mean opening pressure $c_F$ and move the $V_A$ curve to the right. Likewise a lower maximum recruitable fraction $\gamma$ would flatten the $V_A$ curve. Both scenarios indicate a lower compliance and greater pressure required to increase the lung volume in the region of operating pressure.  In certain pathological situations such as ARDS, a sigmoidal representation of $V_A(P_{el})$ with a low compliance region at low $P_{el}$~\cite{Ferreira11,Venegas98,Pereira03, Harris05, Oliveira16} could be captured in the parameterization of $F_{rec}$. 

The viscoelastic properties of pulmonary tissue are represented with a linear Kelvin-Voigt model consisting of scalar compliance $C_{ve}$ and resistance $R_{ve}$, which contributes a viscoelastic pressure component $P_{ve}$ in series with lung elastic recoil $P_{el} $, see Fig~\ref{fig:model}. The sum of these two pressures is dynamic pressure $P_{l,dyn}$ which also equals $P_{pl}-P_A$. 

\subsection*{Respiratory muscle driving pressure}

The pressure $P_{mus}$ describes the effective action of the respiratory muscles driving the model dynamics with $P_{mus}$ negative in the outward direction. We used a sinusoidal function  to describe tidal breathing, with maximum equaling zero at end-expiration:
\begin{equation} 
P_{mus}=A_{mus}\cos (2\pi f t)-A_{mus},
\label{eq:Pmus}
\end{equation}
where $A_{mus}$ is the amplitude of the cosine wave and $f=RR/60$ is the frequency. The wave generates a negative pressure with total magnitude $2A_{mus}$ outward from the body. Though simple, the sinusoidal function can admit time-varying frequency, show dynamics over multiple breaths, is used in artificial ventilation, has compact support on the closed interval $[0,T]$, and has been used in previous modeling studies (see eg.~\cite{Athan00}). More sophisticated functions~\cite{Mecklen98, LeRolle13, Albanese16} can model inhalation and exhalation with different durations or qualitative forms, however the breath-to-breath dynamics displayed in this study can be captured sufficiently with the sinusoidal function. 

\subsection*{Progressive volume loss}\label{sec:loss}

 The complete mechanism of interaction between inefficient inhalation resulting from high chest wall compliance and the progressive nature of lung volume loss and respiratory distress is not fully understood. Clinical X-ray evidence of delayed atelectasis and subsequent acute respiratory distress in otherwise healthy lungs may suggest a process by which a lack of full recruitment during a given breath lowers lung capacity and compliance for the following breath, and continues to an unrecoverable level in the absence of neural modulation or compensatory mechanisms such as sighing. As a first attempt at modeling progressive volume loss, we empirically describe the breath-to-breath evolution of $F_{rec}$ (Eq~(\ref{eq:Frec})) as lung recruitment pressure parameters $c_F$ and $d_F$ increase with PIP and maximum recruitable fraction $\gamma$ decreases with EILV. 

The lung compliance curve shifts slightly with each breath via changes in mean threshold opening pressure based on number of collapsed alveoli. A volume loss associated with derecruited alveoli necessitates an increase in expanded volume of recruited alveoli relative to the radius cubed, with an increased distending pressure proportional to the change in radius. This is illustrated in~\cite{Brown73} using a simple example of expansion of 3 alveoli that double in volume with a 25\% increase in radius; if 1 alveolus closes, the other two radaii must now increase by 35\% to achieve the same overall volume change and the required distending pressure increases proportionally. This proportion applied to $c_F$ and $d_F$ shifts the compliance curve to the right. In this way the compliance decreases approximately proportional to the amount of derecruitment~\cite{Bates09}. If tidal breathing begins on the steepest part of the lung compliance curve, compliance decreases monotonically until eventually tidal breathing occurs on the low compliance tail on the left part of the curve and $V_T\approx 0$. Tidal breathing may begin at a higher position towards the flatter upper part of the curve, in which case compliance will increase slightly with this modification but will again eventually decrease in the manner described above. 

Assuming constant amplitude of the sinusoidal muscle pressure pressure function and no stochasticity, the maximum recruitable fraction of alveoli is achieved at end-inspiration (EI) during steady-state oscillatory breathing and additional fraction will not be recruited under a pressure of this same amplitude in subsequent breaths. The value for $\gamma$ for subsequent breaths is then dependent on $F_{rec}|_{EI}$ and the percentage of alveoli assumed to be permanently collapsed / no longer recruitable, represented by the calculation $\gamma_{next}=\gamma_{current}\dot(1-(\%_{permanent})\cdot(F_{closed}))$ where $F_{closed}$ is fraction closed. If all unrecruited alveoli remain as such, then $F_{rec}|_{EI}$  becomes the new $\gamma$ for the next breath; likewise, if all alveoli remain recruitable, $\gamma=1$ for the duration of the simulation. Note that even for $\gamma=1$, $F_{rec}<1$ for all $P_{el}$ thus causing small changes in $c_F$ and $d_F$ and shifts in the $F_{rec}(P_{el})$ curveregardless of the \% of alveoli permanently closed that still lead to progressive volume loss. The rate at which volume loss progresses depends on where on the $F_{rec}$ curve tidal breathing occurs, and thus both the curve's intrinsic characterizing parameters and extrinsic system variables. 

\section*{Simulation conditions}

\subsection*{Parameterization}\label{sub:physiol}

 The lung curve was parameterized to obtain an approximate dynamic lung compliance $C_A$ of 2.3 ml/cm H$_2$O~\cite{Gerhardt80,Mortola87,Pandit00} calculated as the slope $(V_A|_{EI}-V_A|_{EE})/(P_{el}|_{EI}-P_{el}|_{EE})$ during normal breathing with no interventions. In particular, $k$ was tuned to produce a curve $V_{el}$ between RV and TLC with the calculated slope, and the parameters of $F_{rec}$ produced a curve that is $\approx 1$ for the whole range of normal breathing to represent a nearly fully recruited lung. High $C_w$ for a typical preterm infant was targeted at 8.5 ml/cm H$_2$O~\cite{Gerhardt80,Mortola87} and low $C_w$ about equal to lung compliance. The parameter $d_w$ characterized the approximate dynamic chest wall compliance, which was calculated as the slope $(V_{cw}|_{EI}-V_{cw}|_{EE})/(P_{cw}|_{EI}-P_{cw}|_{EE})$. Parameter values for $R_u$, $R_c$, $R_s$, and $V_c$ were estimated from previously published studies~\cite{Liu98,Athan00,Singh12}. The viscoelastic parameters $C_{ve}$ and $R_{ve}$ were manually tuned to obtain idealized tidal volume and end-expiratory lung volume rather than the magnitude of the hysteresis. 

 FRC is the volume at the resting position of the respiratory system i.e. where $P_{resp}=P_{el}+P_{cw}=0$. The naturally high compliance of the healthy full-term and especially preterm infant (with even steeper $V_{cw}(P_{cw})$) lowers $P_{resp}$ and decreases FRC to about 20\% of vital capacity (VC), compared to at about 35-40\% of VC in the adult~\cite{Agostini86}. A nominal value for FRC for a given set of static compliance curves is obtained by first computing volumes using a vector of physiological pressures [-20...40] cm H$_2$O. Lung and chest recoil pressure vectors are then added in the $P$ direction to obtain $P_{resp}$, and the index where $P_{resp}=0$ is used to determine FRC using either $V_{cw}$ or $V_A$. The lung, chest wall, and respiratory PV compliance curves for both high and low $C_w$ created from Eq~(\ref{eq:Vcw}-\ref{eq:VA}) are given in Fig~\ref{fig:PVcurves}. The value for $P_{el}|_{FRC}$ is then set as the initial condition for solving $dP_{el}/dt$. Note that the lung curve is identical between scenarios so the decreased slope in the low $C_w$ scenario with the same $\nu$ and $V_0$ raises FRC and thus EELV. Decreased lung compliance (flatter $V_A(P_{el})$) resulting from injury, disease, or progressive volume loss further reduces FRC and EELV. In our model we consider chest wall compliance to be either high or low and unchanging for the duration of a simulation, but lung compliance changes depending on breathing conditions.

%%% TABLE 2
\begin{table}[!h]
\begin{adjustwidth}{0in}{1in} % Comment out/remove adjustwidth environment if table fits in text column.
%\centering
\caption{Tuned steady-state and dynamic simulation parameters that remained unchanged during simulations}
{\small
 \begin{tabular}{llccc} 
\toprule
	Parameter 							& Value	& Formula 								& References 				\\ 
\noalign {\smallskip} \hline \hline
\noalign {\smallskip}
TLC [ml]																							&  63		& \textemdash							& \cite{Smith76,Donn98}	\\	
RV [ml]										&  23		& \textemdash							& \cite{Smith76}	\\	
VC[ml]										&  40		&  TLC-RV									& \cite{Smith76,Donn98}	\\	
RR [br/min] 							&  60		& \textemdash							& \cite{Donn98}	\\	
$f$ [br/s]								& 	1		& RR/60										& \textemdash	\\	
$T$ [s]										&  1		& $1/f$										& \textemdash	\\	
$\nu	$							& 0.25	&	\textemdash							& \cite{Donn98,Goldsmith11}	\\	
$V_0$ [ml] 								&  35		& $\nu\cdot$VC+RV	& \textemdash	\\	
$\beta$										&	0.01	&	estimated								& \cite{Hamlington16}  \\
$\gamma$									&		1		&	estimated								& \cite{Hamlington16}  \\
$\alpha$									&	-0.76	&	$\frac{(1+e^{c_F/d_F})\beta-\gamma}{e^{c_F/d_F}}$	& \cite{Hamlington16}  \\
$k$ [1/cm H$_2$O]					&	0.07	&	estimated								& \cite{Ferreira11,Hamlington16}  \\
$c_F$	[cm H$_2$O]					&		0.1	&	estimated								& \cite{Hamlington16}  \\
$d_F$	[cm H$_2$O]					&	0.4		&	estimated								& \cite{Hamlington16}  \\					
$a_w$ [ml]								&  23		& RV											& \cite{Donn98,Goldsmith11}	\\	
$b_w$ [ml]								&  17.3	& $(V_0-RV)/\ln 2$	& \textemdash	\\	
$c_w$ [cm H$_2$O]					&  0		& estimated								& \textemdash	\\	
$a_c$ [ml]								&  0		& \textemdash							& \cite{Liu98}	\\	
$b_c$ [ml]								&  2.5	& $V_{c,max}$							& \cite{Neumann15}		\\
$c_c$ [cm H$_2$O]					&  4.4	& estimated from adult		& \cite{Liu98}	\\	
$d_c$ [cm H$_2$O]					&  4.4	& estimated from adult		& \cite{Liu98}		\\
$K_c$ 										&  0.1		& estimated from adult		& \cite{Athan00}	\\	
$V_{c,max}$ [ml]					&  2.5	& estimated as dead space	& \cite{Donn98,Neumann15}	\\	
$R_{s,m}$ [cm H$_2$O s/L]	&  12		& \textemdash							& \cite{Ratjen92,Singh12}	\\	
$R_{s,d}$ [cm H$_2$O s/L]	&  20	& estimated from adult		& \cite{Athan00}	\\	
$K_s$ 										&  -15	& estimated from adult		& \cite{Athan00}	\\	
$I_u$ [cm H$_2$O s$^2$/L]	&  0.33	& \textemdash							& \cite{Singh12,LeRolle13}\\		
$C_{ve}$ [L / cm H$_2$O] 	& 0.005 & estimated from adult 		& \cite{Athan00}				\\		
$R_{ve}$ [cm H$_2$O s/L] 	& 20		& estimated from adult		& \cite{Athan00}	\\	
\bottomrule
\end{tabular}}
	{\smallskip}
\begin{flushleft} See Table~\ref{tab:glossary} (Glossary) for variable definitions.
\end{flushleft}
\label{tab:SSparams}
\end{adjustwidth}
\end{table}

Table~\ref{tab:SSparams} gives values and formulas / sources for parameters that remain unchanged between simulations. These values as well as the FRC, respiratory pressure amplitude, chest wall compliance, and upper airway resistance parameters in Table~\ref{tab:varparams} that vary between simulation conditions were manually tuned to best obtain the reported aggregate parameters and state outputs as shown in Table~\ref{tab:outputs}. As an example, dynamic lung compliance $C_A$ is not an explicit input into the model, but was determined as described above. For ease of computation and to match the target demographic, we assumed the simulated subject weighed 1 kg. 

%%% TABLE 3
\begin{table}[!h]
\begin{adjustwidth}{-1.4in}{0in} % Comment out/remove adjustwidth environment if table fits in text column.
\centering
\caption{Parameters varying with chest wall compliance and simulation conditions}
{\small
 \begin{tabular}{lcccccc} 
\toprule
\multirow{2}{*}{Parameter}& \multicolumn{2}{c}{High $C_w$}	& \multicolumn{2}{c}{Low $C_w$} 	& \multirow{2}{*}{Formula} 						& \multirow{2}{*}{References} 				\\ 
\noalign {\smallskip}
													&  normal $R_u$ & increased $R_u$		& normal $R_u$ & increased $R_u$	&  						& 									\\ 
\noalign {\smallskip} \hline \hline
\noalign {\smallskip}
FRC [ml] 									&  24.9	& 24.9 											& 28.1	& 28.1										& $P_{el}|_{FRC}+P_{cw}|_{FRC}=0$	& \cite{Smith76,Donn98,Thomas04}	\\	
$A_{mus}$ [cm H$_2$O]			&  1.85	& 3.2 											& 2.78 	& 3.8 										& estimated								& \textemdash	\\	
$d_w$ [cm H$_2$O]					&  0.48	& 0.48 											& 2.4 	& 2.4 										& estimated								& \textemdash	\\	
$R_{u,m}$ [cm H$_2$O s/L]	&  20		& 200 											& 20		& 200 										& \textemdash							& \cite{Mortola87,Singh12}	\\	
$K_u$ [cm H$_2$O s/L]			&  60		& 600 											& 60		& 600 										& estimated from adult		& \cite{Mortola87,Athan00,Singh12}	\\	
\bottomrule
\end{tabular}}
	{\smallskip}
\begin{flushleft} See Table~\ref{tab:glossary} (Glossary) for variable definitions.
\end{flushleft}
\label{tab:varparams}
\end{adjustwidth}
\end{table}
%

%%% TABLE 4
\begin{table}[!ht]
\begin{adjustwidth}{-2.0in}{0in} % Comment out/remove adjustwidth environment if table fits in text column.
\centering
\caption{Aggregate parameters and output states targeted during simulations}
{\small
 \begin{tabular}{lccccccc} 
\toprule
\multirow{2}{*}{Parameter}	& \multirow{2}{*}{Ref. Value} & \multicolumn{2}{c}{High $C_w$}	& \multicolumn{2}{c}{Low $C_w$} 	&  	\multirow{2}{*}{Formula} & \multirow{2}{*}{References} 				\\ 
\noalign {\smallskip}
									&  									& normal $R_u$ & increased $R_u$	& normal $R_u$ & increased $R_u$	&  									&	\\ 
\noalign {\smallskip} \hline \hline
\noalign {\smallskip}
$C_L$ [ml/cm H$_2$O]			& 2.3									& 2.7				& 2.1					& 2.3				&	2.1				&   $\frac{(V_A|_{EI}-V_A|_{EE})}{(P_{el}|_{EI}-P_{el}|_{EE})}$	& \cite{Gerhardt80,Mortola87,Pandit00}	\\	
$C_w$ [ml/cm H$_2$O]			& 8.5									& 9.9				&	16.0				&	2.7			&	3.3				&   $\frac{(V_{cw}|_{EI}-V_{cw}|_{EE})}{(P_{cw}|_{EI}-P_{cw}|_{EE})}$	& \cite{Gerhardt80,Mortola87}	\\	
$C_{rs}$ [ml/cm H$_2$O]		& 1.8									& 2.1				&	1.9				&	1.2			&	1.3				&   $(1/C_L+1/C_w)^{-1}$	& \cite{Mortola87,Neto95}	\\	
$R_{rs}$ [cm H$_2$O s/L] 	& 40									& 34 to 41		&	32 to 223		&	33 to 36		&	32 to 223		&   $R_u+R_c+R_s$ 			& \cite{Goetz01}	\\	
$\dot{V}_E$ [ml/min]			& 360									& 359.2			&	358.2				& 360.0			& 360.0				&   \textemdash				& \cite{Donn98}	\\	
$V_T$ [ml]						& 6									& 5.99			& 5.97				& 6.00 			& 6.00				&  \textemdash					& \cite{Pandit00,Habib03,Schmalisch05} \\		
$\dot{V}_A$ [ml/s]			& $\pm 20$							& -19.4 to 20.8& -16.4 to 28.9 	& -20.3 to 20.4& -17.1 to 26.4	&   \textemdash				& \cite{Pandit00,Habib03}	\\	
$P_{A}$ [cm H$_2$O]			& $\pm 1-2$							&-0.75 to 0.84 & -0.96 to 3.63 	& -0.69 to 0.69& -0.89 to 3.80	&   \textemdash				& \cite{Donn98}	\\	
$P_{el}$ [cm H$_2$O]			& 1 to 6								& 0.9 to 3.6	& 2.7 to 5.9	 	& 1.8 to 4.8	& 2.8 to 6.1		&   \textemdash				& \cite{Pandit00}	\\	
$P_{pl}$ [cm H$_2$O]			& -3 to -6							&-0.6 to -3.8	& 0 to -6.2 		& -1.5 to -4.9 & 0 to 6.4			&   \textemdash				& \cite{Donn98}	\\	
\bottomrule
\end{tabular}}
	{\smallskip}
	\begin{flushleft} EI: end-inspiratory. EE: end-expiratory. See Table~\ref{tab:glossary} (Glossary) for variable definitions.
\end{flushleft}
\label{tab:outputs}
\end{adjustwidth}
\end{table}

\subsection*{Computational procedures}

All simulations proceeded with an initial respiration rate of 60 breaths/min ($f=1$), initial minute ventilation $\dot{V}_E=360$, and initial tidal volume $V_T=6$ ml, with the expectation that tidal volume changes with changes in dynamic lung compliance. The motivation for ths choice was twofold: One, this is consistent with a physiological requirement of constant $\dot{V}_E$ regardless of chest or lung compliance; and two, this allowed for comparison of simulation results originating from similar starting points. Distinct values for $A_{mus}$ were prescribed for each simulation to achieve the initial $\dot{V}_E=360$, see Table~\ref{tab:varparams}. 

 Simulated conditions were chosen to demonstrate the model dynamics with high and low chest wall compliance, under two interventions and two states of permanent alveolar closure. An infant often exhibits compensatory mechanisms such as laryngeal braking (grunting) and increased activity of diaphragm and intercostal muscles~\cite{Smith76} to increase end-expiratory pressure in order to keep EELV above the volume at which alveolar units start to collapse during expiration.  Laryngeal braking is simulated with a 10-fold increase in expiratory upper airway resistance $R_u$. CPAP is simulated with an increase of $P_{ao}$ from 0 to 5 triggered at $F_{rec,EI}=0.9,0.95,0.97$ characterizing volume losses of 10\%, 5\%, and 3\%. Simulations also include assumptions of either no permanently closed alveoli, such that $\gamma=1$ for all time, or10\% permanently closed alveoli per breath, such that $\gamma_{next}=\gamma_{current}\dot(1-0.1(F_{closed}))$.  Each simulation was performed three times: constant $f$; variable $f$ by breath according to $\dot{V}_E=V_{T,ave}\cdot f$ where $V_{T,ave}$ is a moving average of the previous 60 tidal volumes ($\approx 1$ minute of breathing); variable $f$ including a single 20 second apneic event.
	
The system of differential equations~\eqref{eq:states}, together with the constitutive relations~(\ref{eq:Ru}-\ref{eq:Pmus}), were solved using MATLAB R2016b (MathWorks, Natick, MA) with the differential equations solver \verb|ode15s|. Initial conditions were set at physiological values as given in Table~\ref{tab:ICs}. The equations were solved for each new breath using the end conditions from the previous breath as initial conditions. Parameter values as discussed earlier are given in Tables~\ref{tab:varparams} and~\ref{tab:SSparams}. The steady-state stability of the model was analyzed under constant non-oscillatory muscle pressure by examining the eigenvalues of the Jacobian at the nominal parameter set and varying parameters by multiples of 2 and 10. Results of this analysis are found in~\nameref{sec:appendix}.

%%% TABLE 6
\begin{table}[h]
\begin{adjustwidth}{0in}{1in} % Comment out/remove adjustwidth environment if table fits in text column.
\caption{Initial conditions}
{\small
\begin{tabular}{lcccc}	
\toprule 
Initial Condition 			& High $C_w$ 	& Low $C_w$ & Formula 			 				& References		\\ \hline \hline
\noalign {\smallskip}
$\dot{V}(0)$						& 0 					& 0 				&	\textemdash						& \textemdash		\\
$V_c(0)$								& 0.0001			& 0.0001 		&	estimated from adult 	&~\cite{Liu98}	\\
$P_{el}(0)$							& 0.954				& 2.015   	& $P_{el}|_{FRC}$				&~\cite{Smith76}\\ 
$P_{ve}(0)$							& 0 					& 0 				& \textemdash						& \textemdash		\\
\bottomrule
\end{tabular}}
\label{tab:ICs}
\end{adjustwidth}
\end{table}

\section*{Results}

%%% FIGURE 2
\begin{figure}[!h]
\begin{adjustwidth}{-1.53in}{0in} 
\centering
\includegraphics[scale=0.7]{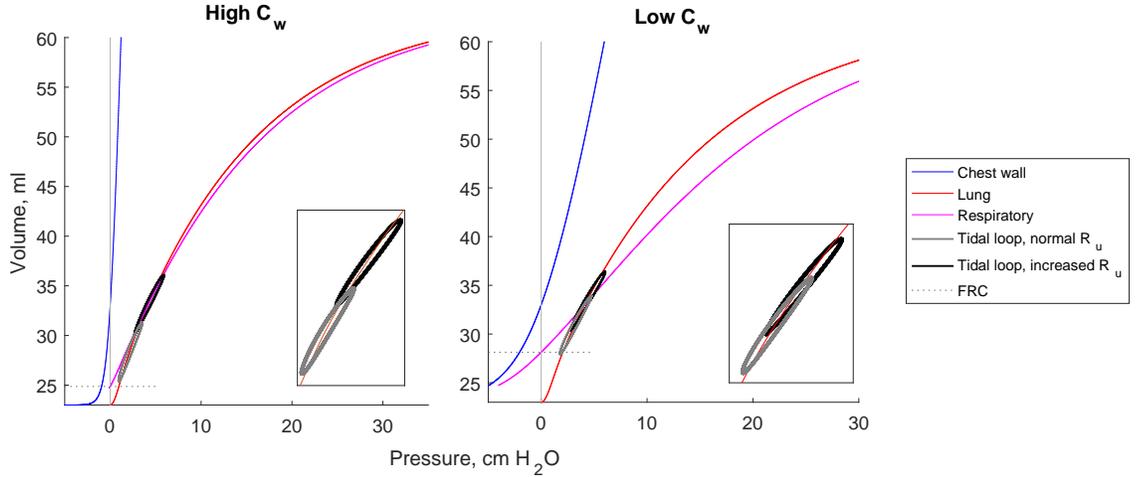}
\caption{{\small {\bf Lung, chest wall, and total respiratory system compliance curves for high $C_w$ (left) and low $C_w$ (right).} Curves are described by Eq~(\ref{eq:Vcw}) and (\ref{eq:VA}) and parameterized using the procedures described in~\nameref{sub:physiol}. Tidal breathing loops with normal $R_u$ (grey) and increased $R_u$ (black) are superimposed for each condition over the lung compliance curve and larger in each inset to display hysteresis.}}
\label{fig:PVcurves}
\end{adjustwidth}
\end{figure}

%%% TABLE 5
\begin{table}[!ht]
\begin{adjustwidth}{-0.9in}{0in} 
\centering
		\caption{Simulations and time to failure (TTF, in hours), defined as 90\% volume loss. }
	{\small
		\begin{tabular}{ccccccc}
\toprule
\multirow{2}{*}{Intervention}& \multirow{2}{*}{$C_w$ condition}	& \multirow{2}{*}{Variable $\gamma$} 	& \multirow{2}{*}{Simulation} & 	\multicolumn{3}{c}{TTF, hours} 						\\ 
\noalign {\smallskip}
									&  										 & 													& 										& constant $f$	& variable $f$ 	& variable $f$ + AE 	\\ 
\noalign {\smallskip} \hline \hline
\noalign {\smallskip}
    \multirow{4}{*}{None} 					& {\multirow{2}{*}{Low}} 		& No 										& 1 					& 2.49 				&	2.53 & 2.51\\
		\cmidrule{3-7}          				&	       						& Yes										& 2 					&	2.25 				&	2.28 & 2.26 \\
		\cmidrule{2-7}         					& {\multirow{2}{*}{High}} 	& No 			 							& 3 					&	0.30 				&	0.32 & 0.30 \\
		\cmidrule{3-7}          				&       							& Yes										& 4 					&	0.27 				& 0.29 & 0.27 \\
		\midrule
    \multirow{4}{*}{Increased $R_u$}& {\multirow{2}{*}{Low}} 		& No  										& 5 					&	24.7 				&	24.7 & 24.5\\
		\cmidrule{3-7}          				&	       						& Yes										& 6 					&	22.2 				&	22.2 & 21.9 \\
		\cmidrule{2-7}         					& {\multirow{2}{*}{High}} 	& No 			 							& 7 					&	18.5 				&	18.4 & 17.3\\
		\cmidrule{3-7}          				&       							& Yes										& 8 					&	16.6 				& 16.5 & 15.0 \\
    \midrule
     \multirow{4}{*}{CPAP, 10\% loss}	& {\multirow{2}{*}{Low}}	& No  										& 9 					&	2.56 				&	2.61 & 2.60 \\
		\cmidrule{3-7}          				&	       						& Yes										& 10 					&	2.32 				&	2.36 & 2.34 \\
		\cmidrule{2-7}         					& {\multirow{2}{*}{Yes}} 	& No 			 							& 11 					&	0.83 				&	0.79 & 0.76 \\
		\cmidrule{3-7}          				&       							& Yes										& 12 					&	0.83 				& 0.80 & 0.77 \\
		\midrule
		CPAP, 5\% loss									& {\multirow{2}{*}{High}} 	& {\multirow{2}{*}{No}} 		& 13 					&	2.94 				& 2.50 & 2.46 \\
		\cmidrule{1-1} \cmidrule{4-7} 
		CPAP, 3\% loss									&										&									&	14				  & 8.57  				& 6.89 & 6.83 \\
		\bottomrule
  \end{tabular}}
	{\smallskip}
\begin{flushleft} Increased $R_u$: A 10-fold increase in $R_u$ was applied during expiration. CPAP: Simulated administration of $P_{ao}=5$ occurred when recruited fraction was down 10\%, then again at 5\% and 3\% with constant $\gamma$. AE: A single 20 second apneic event occurred at the 2 minute mark of the simulation.
\end{flushleft}
\label{tab:simulations}
\end{adjustwidth}
\end{table}

Parameterized static compliance curves for $V_{cw}(P_{cw})$ and $V_A(P_{el})$ are shown in Fig~\ref{fig:PVcurves} for high $C_w$ (left) and low $C_w$ (right). The hysteretic tidal breathing loops are superimposed on the curve $V_A(P_{el})$ for normal $R_u$ in black and increased $R_u$ in grey. Hysteresis is caused in the model by the viscoelastic parameters $C_{ve}$ and $R_{ve}$, which were tuned to maintain appropriately valued lung volume outputs.

\begin{figure}[!h]
\begin{adjustwidth}{-2in}{-0in} 
\centering
 \includegraphics[scale=0.7]{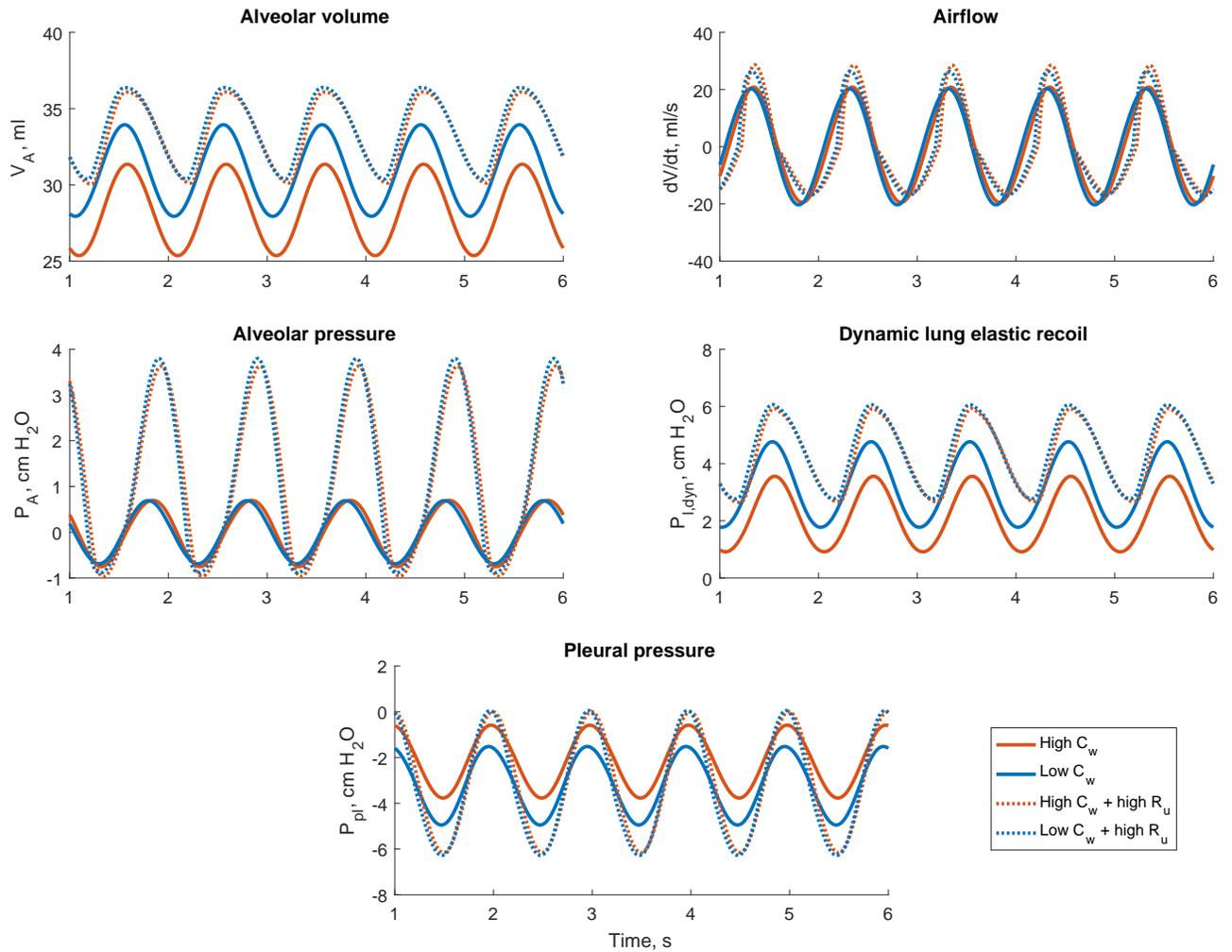}
\caption{{\small {\bf Simulated periodic steady-state tracings of five breaths.} Depicted are alveolar volume, airflow, alveolar pressure, dynamic elastic lung recoil, and pleural pressure, under high and low $C_w$ conditions, with normal vs. high $R_u$.}}
\label{fig:SSstates}
\end{adjustwidth}
\end{figure}

Fig~\ref{fig:SSstates} shows the impact of high vs low $C_w$ and normal vs. high $R_u$ on the five states $P_A, P_{l,dyn}, P_{pl}, V_A$, and $\dot{V}$. Increased $R_u$ increases $P_A$ almost threefold, but $C_w$ has very little impact. However, decreased $C_w$ increases $P_{l,dyn}$ significantly, effectively raising it higher on the lung PV curve. Increasing $R_u$ even higher increases $P_{l,dyn}$ but there is no difference with respect to $C_w$. The opposite appears to occur with $P_{pl}$ dynamics, in that decreasing $C_w$ makes $P_{pl}$ more negative (``increasing'' the magnitude of the pressure) and increasing $R_u$ strengthens that effect. Low $C_w$ and subsequently high $R_u$ increase $V_A$, mimicking the effect for $P_{l,dyn}$. High $R_u$ shifts $\dot{V}$ by ~5 ml/s, with airflow more restricted during expiration. Tabulated magnitudes of the steady states are gives in Table~\ref{tab:outputs}. These results compare favorably to the record reported in Abbasi et al~\cite{Abbasi90} in which esophageal (pleural) pressure, airflow, and tidal volume were approximately -2 to -6 cm H$_2$O, -30 to 30 ml/s, and 8 ml, respectively.

%%% FIGURE 3

 Table~\ref{tab:simulations} presents the 14 simulations and their time to failure, defined for this study as 90\% volume loss. Dynamics were comparable between simulations with the major difference being the timing, therefore only representative or significant results are presented in figures. Our model consistently indicates a faster loss of end-expiratory lung volume in all simulations with high $C_w$ compared to the same with low $C_w$. Variable $f$ did not significantly change TTF except in the case of CPAP administered at 3\% loss (S14), with TTF shortened by almost 2 hours. Adding a single 20 second apnea shortened the TTF by ~1-4 minutes in the shortest simulations but by over an hour under high $C_w$ and increased $R_u$ (S8).
 
\begin{figure}[!h]
%\begin{adjustwidth}{-.1in}{0in} 
\centering
\centerline{\includegraphics[scale=0.7]{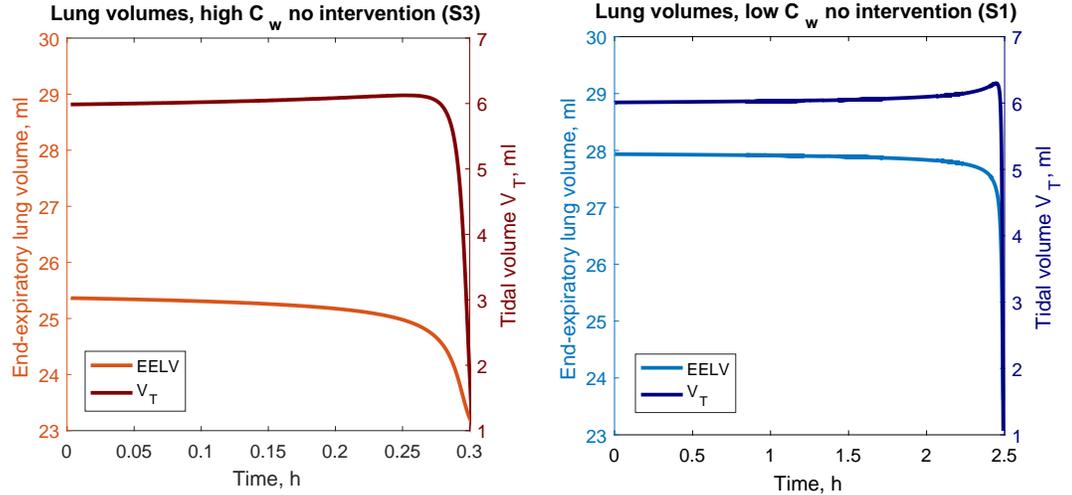}}
\caption{{\small {\bf Breath-to-breath volumes.} End-expiratory lung volume (left y-axis) and tidal volume (right y-axis) under high and low $C_w$ conditions, no interventions.}}
\label{fig:lungvolumes}
%\end{adjustwidth}
\end{figure}

The breath-to-breath change in EELV and $V_T$ under high and low $C_w$ conditions with no interventions are given in Fig~\ref{fig:lungvolumes} (Simulations 1 and 3). The high $C_w$ simulation reaches accelerated loss of volume and eventually failure at ~0.3 hours, much more quickly than the low $C_w$ at ~2.5 hours. This depicts a possible scenario in which lung volume loss and failure may appear to onset suddenly after a long period of apparent steady conditions.

%%% FIGURE 4

\begin{figure}[!h]
%\begin{adjustwidth}{-1.1in}{0in} 
\centering
\centerline{\includegraphics[scale=0.65]{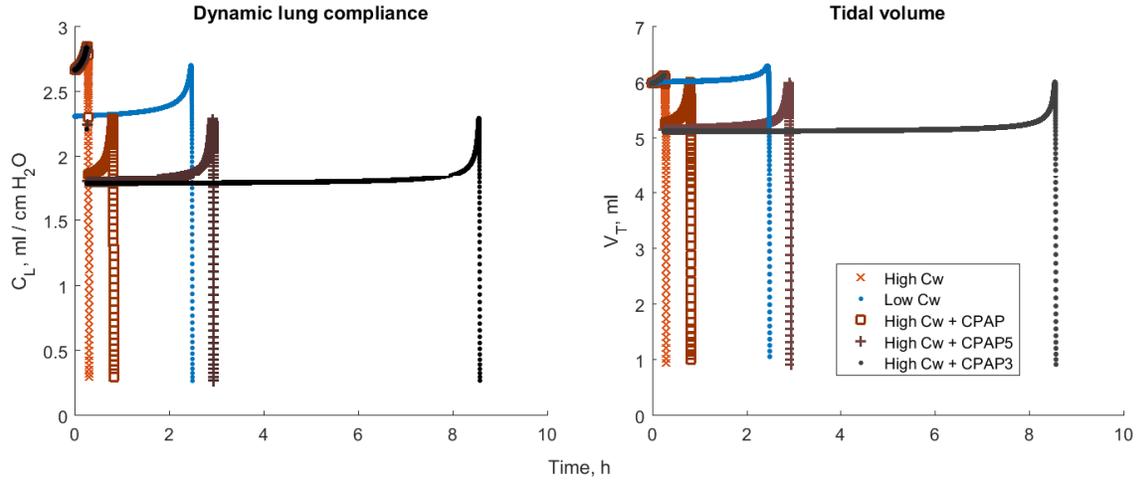}}
\caption{{\small {\bf Breath-to-breath dynamic lung compliance and tidal volume}. Depicted are high and low $C_w$ conditions, with simulated CPAP triggered in the high $C_w$ condition when recruited fraction dropped 10\%, 5\%, and 3\%.}}
\label{fig:Cdyn}
%\end{adjustwidth}
\end{figure}

Fig~\ref{fig:Cdyn} shows changes in dynamic lung compliance and tidal volume with high and low $C_w$ without changes in $\gamma$, then adding CPAP to the high $C_w$ condition at three different levels (c.f. Table~\ref{tab:simulations}, simulations 1,3,11,13-14). CPAP was simulated by an increase in mouth pressure $P_{ao}$ to 5 cm H$_2$O when $F_{rec,max}<0.9$, which happened when the lung volumes were already decreasing quickly towards failure. However, CPAP triggering at $F_{rec,max}<0.95$ and 0.97 gained ~3 and ~9 hours of time, respectively. Note that regardless of timing, the  administration of CPAP is correlated with reduced tidal volume (see also ~\cite{Waugh99}. Increasing $P_{ao}$ moves the resulting PV loop higher up on the lung compliance curve but does not change the nature of the curve, thus eventually the influence of high $C_w$ on dynamics induces the same lung volume loss without other mitigating actions. 

%%% FIGURE 5

\section*{Discussion}

In summary, we have developed a lumped-parameter respiratory mechanics model tuned with parameters specific to the extremely preterm infant weighing 1 kg. The model includes a novel representation of derecruitment based on alveolar pressure and volume expansion compensating for collapsed alveoli. Model simulations suggest conditions under which volume loss may result more quickly from higher vs lower chest wall compliance in the preterm infant, indicating the plausibility of dynamics underlying the symptoms observed clinically. Given the fragile nature of this population, it is extremely difficult to obtain non-pathological parameter or state output values for a healthy or surfactant-treated infant during spontaneous breathing, and even more so to obtain time series for model validation and eventual parameter estimation. The much earlier study by Abbasi and Bhutani~\cite{Abbasi90} and a later one by Pandit et al~\cite{Pandit00} gave the best insight into the respiratory dynamics of an extremely preterm infant, making these the standard against which our results were qualitatively validated. We therefore claim that this effort is a ``proof of concept'' that will be further explored in future investigations using pressure and airflow time series data in a parameter estimation / optimization procedure to characterize parameter values specific to a particular patient dataset. Additional model modifications will allow for hypothesis generation for future data testing and data collection. 
  
As mentioned briefly in~\nameref{sec:compliances}, recruitment/decrecruitment may have a time component~\cite{Bates02,Albert09}, in that  the time it takes for an airway or alveolus to open may be a function of how far away its pressure is from its critical opening pressure. Earlier studies have developed models that incorporate opening and closing pressures for individual alveoli, contributing to the aggregate difference in inflation and deflation limbs of the hysteretic PV curve~\cite{Crotti01,Markhorst04}. These previous studies considered recruitment resulting from one or two hyperinflations but not long-term derecruitment. In our model breath-to-breath derecruitment is manifested as the change of the lung compliance curve during normal spontaneous breathing as described in~\nameref{sec:loss}, and the hysteresis found in the tidal breathing loop is accounted for by the viscoelastic component of the system of differential equations. It is clear from Table~\ref{tab:simulations} that time to failure shortens if an assumption is made about a non-zero percentage of alveoli permanently closing and being unavailable for recruitment. As a topic for further study, the pulmonary tissue may be modeled by more complex Voigt-Maxwell models within the "electrical analog" model or other non-electrical analog representations (e.g. those described in~\cite{Bates02}). While such a modification may affect the overall trends in observed states such as EELV, the differential impact between high and low $C_w$ would be expected to remain.

The noninvasive ventilatory intervention CPAP shifts the tidal volume loop to a higher position on the lung compliance curve, operating with a higher EELV and end-expiratory lung elastic recoil. Our model suggests that the timing of administration of CPAP and the permanent closure or injury state of alveoli may impact its effectiveness. In our first simulation with simulated CPAP triggered at 10\% volume loss, the recruited volume fraction does not recover fully to 1 and the use of CPAP only gains about a half hour of breathing before failure. However, CPAP starting at 5\% and 3\% loss gained 3 and 9 hours of time, respectively. This magnitude of loss may not be symptomatic at this point but would benefit from pressure support to avoid the quick descent to failure. These results are reported for the case with all fully recruitable alveoli. In the case of 10\% permanent collapse of closed alveoli at each breath and subsequent breath-by-breath decrease in $\gamma$, the function $F_{rec}$ can never reach 1 (full recruitment) for the duration of breathing, tidal breathing occurs on a lower lung compliance curve, and CPAP cannot recover the full volume loss in subsequent breaths. Results in Table~\ref{tab:simulations} indicate that time to failure is ~10\% faster with the permanent collapse. These simulated loss and collapse percentages were arbitrarily chosen to demonstrate the capabilities of the model and possible influences on breathing dynamics, but more investigation into actual loss values would add to the model's usefulness. Starting from a lower $C_w$ appears to be the optimal condition presented here as hypothesized. 

Prolonged shallow breathing has been associated with increased surface tension and decreased surface area that further hinders breathing~\cite{Williams66}. We safely assume in our model that derecruitment is a continuous process that will eventually induce loss of lung volume if left uncompensated~\cite{Mead59,Ferris60}. In a healthy lung in the absence of fatigue, permanent alveolar collapse (due to injury or disease), and/or high chest wall compliance, this process is on a much longer time scale than the natural compensation mechanisms that compensate for and recoup volume loss (such as grunting in the infant). One such mechanism is spontaneous deep breathing, or ``sighing'', which may help prevent atelectasis~\cite{Bartlett71,Duggan05,Qureshi08} by re-opening air spaces that collapse naturally under tidal breathing~\cite{Bendixen63} via increased pressure and surfactant activation and possibly affect neurorespiratory control. Sighing occurs more frequently and at relatively larger magnitude in the infant vs adults~\cite{Davis94}. A natural extension of our model would be incorporating the restorative actions of sighing and testing the hypothesis that spontaneous deep breaths mitigate or reverse volume loss. 

 Several features of the physiology of preterm infants are not currently addressed in this model but should be considered in future model enhancements for further investigations. Preterm infants commonly exhibit diaphragm weakness and dysfunction and paradoxical breathing.  While a sinusoidal waveform is used in the clinic under some mechanical ventilation protocols, the sinusoidal pressure function used here is an elementary representation for spontaneous breathing and does not capture dynamics related to diaphragm dysfunction or possible expiratory flow limitation. Modifications reflecting such dynamics may include adjustments to the pressure amplitude, varying fractions of time spent in inspiration vs expiration, and the use of a model that combines functional forms such as polynomials or exponentials (see e.g.~\cite{Mecklen98}). Components that differentiate between abdominal and rib cage movements (see e.g.~\cite{Primiano82}) may model the paradoxical chest movement. 

Another limitation of this model is the absence of any feedback mechanisms compensating for loss of volume. More sophisticated models of central pattern generators have been developed in conjunction with simple lung mechanics~\cite{Jallon09,Diekman17} that could potentially be incorporated with ours. A chemoreflex model, see for example~\cite{Chiari95,Diekman17}, may also augment our model. Despite these limitation, we expect that the timing of dynamics of individual simulations may change with model enhancements but that time to failure would still be extended under low chest wall compliance conditions as observed in this study.

\section*{Conclusion}

Respiratory mechanics models have been investigated for several years and many formulations exist; the challenge to be appreciated is the customization to the preterm infant with significantly different physiological features than adults and even term infants. Hence future model modifications must always keep this at the forefront of any investigation. The lumped-parameter respiratory mechanics model developed in this study will be used in future studies with data currently being collected in the NICU to estimate patient-specific parameters, which may shed light on factors influencing volume loss dynamics. This process may help generate hypotheses about predicting volume loss and recovery to motivate future data collection strategies. Our hope is that these investigations lead to a chest-stiffening treatment that can target an infant's specific physiological characteristics and prevent volume loss in this vulnerable population. 

\section*{Acknowledgments}
This research was supported in part by the Atlantic Pediatric Device Consortium via FDA grant 5P50FD004193-07 (H. Rozycki, L. Ellwein, M. Brandes) and the VCU College of Humanities and Sciences Faculty Research Council (L. Ellwein, L. Linkous). 

\section*{Appendix}\label{sec:appendix}

%\appendix

We analyze the inherent stability of the model under constant non-oscillatory muscle pressure by examining the eigenvalues of the Jacobian at the nominal parameter set and varying parameters by multiples of 2 and 10. To obtain steady-states, $P_{mus}$ is set at a constant called $P_{mus,C}$ and the system of ODE's is set to equal 0:
\begin{eqnarray*}
0&=&\frac{1}{I_u}\left(P_{ao}-P_u-R_u\dot{V}\right)\\
0&=&\dot{V}-\dot{V}_A\\
0&=&\frac{\dot{V}_A}{C_A}\\
0&=&\frac{\dot{V}_A-(P_{ve}/R_{ve})}{C_{ve}}
\end{eqnarray*}
It follows that $\dot{V}=\dot{V}_C=\dot{V}_A=0$ and $P_u=P_{ve}=P_c=P_A=0$. Given the relations $P_{el}=f(V_A)$, $P_{tm}=f(V_c)$, and $P_{cw}=f(V_{cw})$,  we have 3 equations with 6 unknowns $(V_A,V_c,V_{cw},P_{el},P_{tm},P_{cw}$). Three additional equations come from incorporating loop summations such that $P_{el}(V_A)=P_{tm}(V_c)$, $P_{tm}=-(P_{cw}+P_{mus,C})$, and $V=V_{cw}=V_A+V_c$. Noting that $P_{el}$ is defined implicitly by Eq.~(10) and using the compliance curve functions
\begin{eqnarray*}
P_{cw}&=&c_w+d_w\ln \left(e^{(V_A+V_c-a_w)/b_w}-1\right)\\
P_{tm}&=&c_c-d_c \ln \left(\frac{V_{c,max}}{V_c}-1\right)
\end{eqnarray*}
 we obtain two equations
\begin{eqnarray*}
0&=&P_{cw}(V_A(P_{el}),V_c)+P_{mus,C}+P_{tm}(V_c)\\
0&=&P_{cw}(V_A(P_{el}),V_c)+P_{mus,C}+P_{el}
\end{eqnarray*}
that are solved numerically for two unknowns $P_{el},V_c$ for each modified parameter set using an iterative algorithm. $P_{el}$ From these steady-state values we can calculate the remaining state variables.

In order to linearize the system and determine asympototic behavior we rewrite the system in terms of the state variables $\dot{V},V_c,P_{el},P_{ve}$ using previously described relationships:
\begin{eqnarray*}
\ddot{V}&=&\frac{1}{I_u}\left(P_{ao}-R_c(V_c)\dot{V}-P_{tm}(V_c)-P_{cw}(V_A(P_{el})+V_c)-P_{mus}-R_u(\dot{V})\dot{V}\right)\\
\dot{V}_c&=&\dot{V}-\frac{P_{tm}(V_c)-P_{el}-P_{ve}}{R_s(V_A(P_{el}))}\\
\dot{P}_{el}&=&\frac{1}{C_A(P_{el})}\left[\frac{P_{tm}(V_c)-P_{el}-P_{ve}}{R_s(V_A(P_{el}))}\right]\\
\dot{P}_{ve}&=&\frac{1}{C_{ve}}\left[\frac{P_{tm}(V_c)-P_{el}-P_{ve}}{R_s(V_A(P_{el}))}-\frac{P_{ve}}{R_{ve}}\right]
\end{eqnarray*}
where the quantity $\frac{P_{tm}(V_c)-P_{el}-P_{ve}}{R_s(V_A(P_{el}))}=\dot{V}_A.$ The Jacobian
\[ J=
\begin{bmatrix}
\ffrac{\partial\ddot{V}}{\partial\dot{V}} & \ffrac{\partial\ddot{V}}{\partial V_c} & \ffrac{\partial\ddot{V}}{\partial P_{el}} & \ffrac{\partial\ddot{V}}{\partial P_{ve}} \\
\ffrac{\partial\dot{V}_c}{\partial\dot{V}} & \ffrac{\partial\dot{V}_c}{\partial V_c} & \ffrac{\partial\dot{V}_c}{\partial P_{el}} & \ffrac{\partial\dot{V}_c}{\partial P_{ve}} \\
\ffrac{\partial\dot{P}_{el}}{\partial\dot{V}} & \ffrac{\partial\dot{P}_{el}}{\partial V_c} & \ffrac{\partial\dot{P}_{el}}{\partial P_{el}} & \ffrac{\partial\dot{P}_{el}}{\partial P_{ve}} \\
\ffrac{\partial\dot{P}_{ve}}{\partial\dot{V}} & \ffrac{\partial\dot{P}_{ve}}{\partial V_c} & \ffrac{\partial\dot{P}_{ve}}{\partial P_{el}} & \ffrac{\partial\dot{P}_{ve}}{\partial P_{ve}} \\
\end{bmatrix}
\]
was found using symbolic computation then used numerically to calculate eigenvalues. All parameter variations gave stable solutions except $c_c, d_c$, and $V_{c,max}$. However, varying these parameters by 2 and 10 is not actually physiological and the instability comes from these modifications making the parameter values inconsistent with the rest of the system. The parameters $c_c$ and $d_c$ must be equal with our calculations, and since $c_c$ is a normal operating pressure, it must stay within a narrow range. Likewise, $V_{c,max}$ is equivalent to dead space so not only must it also be within a narrow range but if it varies too much with respect to other parameters and states then the equations will be inconsistent. It is therefore reasonable to state that the steady-state of this model system is inherently stable for the physiological parameter ranges used here.

\nolinenumbers

% Either type in your references using
% \begin{thebibliography}{}
% Text
% \end{thebibliography}
%
% or
%
% Compile your BiBTeX database using our plos2015.bst
% style file and paste the contents of your .bbl file
% here. See http://journals.plos.org/plosone/s/latex for 

%\bibliography{MathBiosci_PLOS}

\end{document}